\documentclass{llncs}

\usepackage[utf8]{inputenc}

\usepackage{url}
\usepackage{graphicx}
\usepackage{xcolor}
\usepackage[font=footnotesize, skip=10pt, hangindent=10pt]{caption}
\usepackage{subfigure}
\usepackage{booktabs}
\usepackage{listings}
\usepackage{cite}

\usepackage{amsmath}
\usepackage{amssymb}

\usepackage{amsthm}

\newcommand{\refSec}[1]{Sect.~\ref{#1}}
\newcommand{\refFig}[1]{Fig.~\ref{#1}}
\newcommand{\refTable}[1]{Tab.~\ref{#1}}

\begin{document}

\title{Unlinkable content playbacks in a\\ multiparty DRM system (full version)\thanks{This is the full version of the paper that appeared in the Proceedings of The 27th Annual IFIP WG 11.3 Working Conference on Data and Applications Security and Privacy (DBSec '13). The final publication is available at link.springer.com}} 
\author{Ronald Petrlic (ronald.petrlic@upb.de)\\ Stephan Sekula (sekula@live.upb.de)}
\institute{University of Paderborn\\ 33098 Paderborn, Germany}

\maketitle

\begin{abstract}

We present a solution to the problem of privacy invasion in a multiparty digital rights management scheme. (Roaming) users buy content licenses from a content provider and execute it at any nearby content distributor. Our approach, which does not need any trusted third party---in contrast to most related work on privacy-preserving DRM---is based on a re-encryption scheme that runs on any mobile Android device. Only a minor security-critical part needs to be performed on the device's smartcard which could, for instance, be a SIM card.

\end{abstract}

\section{Introduction} \label{introduction}

Mobile users are used to access digital content provided in the cloud from anywhere in the world today. Music streaming services like \emph{Spotify} enjoy great popularity among users. The lack of bulky storage on mobile devices (mobile phones, tablets, etc.) is compensated for by such services by streaming the content (music and films) to the users's devices. Content is downloaded on demand and can be used only during playback (or is cached temporarily). Thus, paying users are able to access huge amounts of content. There also exist certain price models that allow the playback of content only for a certain number of times, until a specific day (e.g., movie rentals), etc.

In a ``simple'' digital rights management (DRM) architecture, there would be only two parties involved: a user who wants to get access to content and the content provider that either grants or denies access---depending on the payment status. However, in more advanced and realistic scenarios, we would have a content provider that sells licenses---granting certain rights for content playback---to users and there would be a number of content distributors that provide the (protected) content. This allows for a better scaling of the service as users can access content from those content distributors that are closest or that provide best service at the moment. This bears advantages for roaming users as they can choose local distributors to access the content from. For the content distributor to be able to decide whether the user is allowed to playback the content, the user first needs to present the purchased license. Such scenarios, consisting of content providers and distributors, are called \emph{multiparty} DRM systems in the literature. 

A drawback of most of today's DRM systems is that content providers/distributors are able to build content usage profiles of their users as they learn which user plays back certain content at a certain time, etc. This is the point where we contribute with this paper. We suggest a privacy-preserving multiparty DRM system. In such a system, users are able to anonymously buy content and anonymously playback the content. Moreover, neither the content providers nor the content distributors are able to link content playbacks to each other and thus are not able to build usage profiles under a pseudonym---as the past has shown that even profiles under a pseudonym, assumed to be unrelatable to users, can be related to users given external information and thus, inverting user privacy again \cite{netflix}. One major advantage of our approach compared to related work on privacy-preserving DRM is that we do not need a trusted third party (TTP) which needs to check licenses before allowing content executions. 

The paper is structured as follows. Related work is covered in \refSec{related_work} and the preliminaries in \refSec{preliminaries}. In \refSec{system_model} we present the system model of our multiparty DRM scenario and present the corresponding requirements. Our proposed privacy-preserving multiparty DRM system is presented in \refSec{concept}. We discuss and evaluate our proposed concept in \refSec{evaluation} before we conclude in \refSec{conclusion_outlook}. 


\section{Related Work} \label{related_work}

In this section we first give a short overview of related work on the building blocks of our proposed concept. Then we investigate state-of-the-art research of privacy-preserving DRM. 

\subsection{Proxy Re-Encryption}

\emph{Proxy re-encryption} allows a proxy to transform an encrypted message under $A$'s public key into another encrypted message under $B$'s public key---without seeing the message in plain text. For this, a re-encryption key $rk_{A \rightarrow B}$ is used. The proxy does not need the private key of $A$ to decrypt the message and encrypt it again under $B$'s public key. \textsc{Ateniese et al.} \cite{ateniese_improved} introduce several \emph{unidirectional} proxy re-encryption schemes---one of them is covered in \refSec{preliminaries}. 

\subsection{Anonymous Payments} \label{payment}

The anonymous payment scheme used in this paper has been introduced by \textsc{Tewari et al.} \cite{Tewari1998}. Analyses of this method have proven it to be more suited for our application in comparison to the frequently cited scheme by \textsc{Chaum} \cite{chaum_cash} in \cite{architecture}. This is because of the chosen system being more flexible. Another reason behind choosing this protocol is that the \emph{Point of Sales} (POS) devices used provide another layer of anonymity for the user since these device serve as a proxy between the user and the payee. Contrasting the basic version \cite{Tewari1998}, however, we provide extensions such as the payment of change in case the user does not have the right amount of money in his/her wallet. The anonymous payment scheme does not allow any party to get to know which content has been purchased if the user makes legitimate payments only. If the user tries to defraud some party, however, his/her identity can be unveiled so that he/she can be held accountable.

\subsection{Privacy-preserving DRM}

In \cite{multilevel}, \textsc{Mishra et al.} propose a scenario where a content owner provides its (encrypted) content to users via a number of different (local) content distributors---which is similar to our scenario. Employing this scheme, users can buy licenses for content from a license server, acting as trusted third party. Once a license is bought, the user gets in possession of the decryption key which allows him/her to access the content as often as desired. Differentiated license models are not intended in their approach---however, if license enforcement additionally took place on the client-side, differentiated license models could be implemented. As content download and license buying are done anonymously, none of the parties can build profiles of users' interest in content.\\

\textsc{Win et al.} \cite{nottp} present a privacy-preserving DRM scheme for two- and multiparty scenarios without needing a TTP. A user anonymously requests a token set from the content owner that allows anonymous purchase of content licenses from content providers. A drawback is that content providers are able to build usage profiles of content executions under a pseudonym. 

\textsc{Petrlic et al.} \cite{petrlic1} present a DRM scenario that allows users to anonymously buy software from any software provider and execute it at any computing center within the cloud. The users' permission to execute the software is checked before every single execution. Their solution is resistant against profile building. The authors suggest employing a software re-encryption scheme that is based on secret sharing and homomorphic encryption to achieve unlinkability of software executions towards the computing center. Their software re-encryption scheme is rather complex and implies a huge communication overhead. The approach is extended in~\cite{proxy} by employing an adapted version of proxy re-encryption \cite{ateniese_improved}. The scheme makes explicit use of a service provider as a TTP. 

The approach towards privacy-preserving DRM by \textsc{Joshi et al.} \cite{practical} also requires a TTP for license checking before content execution. It makes use of a number of cryptographic primitives such as proxy re-encryption, ring signatures and an anonymous recipient scheme to provide unlinkability of content executions. The scheme's advantage is the reduced computation and communication overhead compared to the approaches above.

\section{Preliminaries} \label{preliminaries}

\begin{definition}{Unidirectional Proxy Re-Encryption} \cite{ateniese_improved}
\label{re-enc}

A unidirectional proxy re-encryption scheme is a tuple of polynomial time algorithms ($KG$, $RG$, $E_{1}$, $E_{2}$, $R$, $D$). The randomly chosen $g \in \mathbb G_{1}$ and $Z = e(g,g) \in \mathbb G_{2}$ constitute the global system parameters.

\begin{itemize}

\item \emph{Key Generation} (KG): KG generates a public key $pk_{a} = (Z^{a_{1}}, g^{a_{2}})$ and a private key $sk_{a} = (a_{1}, a_{2})$ for $A$, where $a_{1}, a_{2} \in \mathbb Z_{q}$ are chosen randomly. 

\item \emph{Re-Encryption Key Generation} (RG): $A$ delegates to $B$ by publishing $rk_{A \rightarrow B} = g^{a_{1} b_{2}} \in \mathbb G_{1}$. For that purpose, $A$ needs $B$'s public value $g^{b_{2}}$. 

\item \emph{First-Level Encryption} ($E_{1}$): $m \in \mathbb G_{2}$ is encrypted under $Z^{a_{1}} \in pk_{a}$ by outputting $c = (Z^{a_{1}k}, mZ^{k})$, where $k \in \mathbb Z_{q}^{*}$ is chosen randomly. Note that $Z^{a_{1}k} = e(g^{a_{1}}, g^{k})$. The message can be decrypted by the holder of $a_{1} \in sk_{a}$. 

\item \emph{Second-Level Encryption} ($E_{2}$): $m \in \mathbb G_{2}$ is encrypted under $pk_{a}$ by outputting $c = (g^{k}, mZ^{a_{1}k})$. Second-level ciphertexts can be transformed to first-level ciphertexts by re-encrypting them. 

\item \emph{Re-Encryption} (R): A second-level ciphertext for $A$ can be changed into a first-level ciphertext for $B$ with $rk_{A \rightarrow B} = g^{a_{1}b_{2}}$. From $c = (g^{k}, mZ^{a_{1}k})$, compute $e(g^{k},g^{a_{1}b_{2}}) = Z^{b_{2}a_{1}k}$ and output $c = (Z^{b_{2}a_{1}k}, mZ^{a_{1}k})$. 

\item \emph{Decryption} (D): 

\begin{itemize}
\item A (re-encrypted) first-level ciphertext $c = (\alpha, \beta)$ (for $B$) is decrypted with private key $b_{2} \in sk_{B}$ by computing $m = \frac{\beta}{\alpha^{1/b_{2}}}$. 
\item A first-level ciphertext $c = (\alpha, \beta)$ (for $A$)---computed using $E_{1}$---is decrypted with private key $a_{1} \in sk_{A}$ by computing $m = \frac{\beta}{\alpha^{1/a_{1}}}$.
\end{itemize}

\end{itemize}

\end{definition}


\section{System Model} \label{system_model}

\begin{figure*}[ht]
	\centering
	\includegraphics[width=0.75\textwidth]{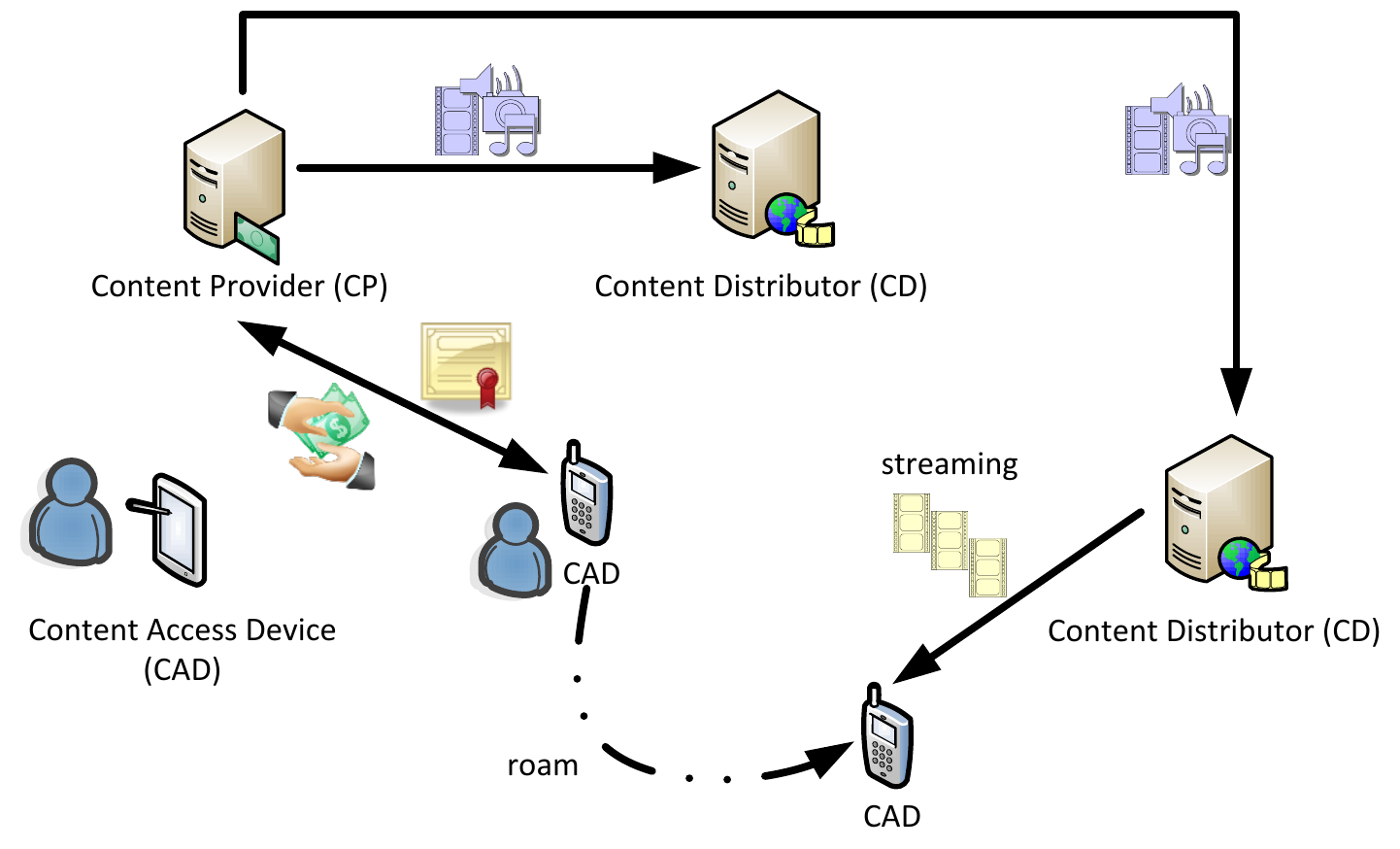}
	\caption{System model of a multiparty DRM scenario.}
	\label{fig:system_model}
\end{figure*}

The system model is shown in \refFig{fig:system_model} on a high abstraction level. We are dealing with a multiparty DRM scenario that involves \emph{content providers} (CPs), \emph{content distributors} (CDs), and \emph{users}. The focus, as shown in the figure, is on mobile users with different \emph{content access devices} (CADs) accessing content. However, the scenario is not limited to mobile devices. As devices have different hardware trust anchors---e.g., smartphones typically are equipped with SIM cards, tablet computers have \emph{trusted platform modules} (TPMs), etc.---we subsume those trust anchors under the term \emph{smartcards} in the following.\footnote{Note that SIM cards are smartcards and TPMs can be seen as a special form of smartcards as well.}

The CP takes the role of, e.g., a film studio or music label that produces content. Users interact with the CP by buying a license that allows playback of the content---under certain terms that are mediated. The user's smartcard is used to check whether the user is still allowed to access the content. Then, a nearby CD is contacted and the CD streams the content to the user. The CD can have contracts with different CPs, which allows the user to access content by different CPs from a single source---as it is the case with state-of-the-art streaming servers as well. The CD might get paid for providing its services by the CPs (or even the users). We do not cover this aspect in the paper at hand.

We assume that the CPs and CDs are \emph{honest-but-curious}, i.e., they follow the protocol, as specified in \refSec{concept}, but they try to find out as much as possible to track users. Users can be assumed as \emph{active adversaries}, meaning that they try to break the protocol to be able to execute content without having a license.

Our protocol is not based on any TTP that would be needed to check licenses.



\subsection{DRM requirements} \label{requirements}

In a multiparty DRM scenario, there are a number of stakeholders with diverse---sometimes even contrary---requirements. In our scenario, we can identify the \emph{content provider}, \emph{content distributor}\footnote{As stated above, we do not investigate the CD as a stakeholder in detail---we can assume that it gets paid by the CP or the user for its services.}, and the \emph{user} as main stakeholders. The requirements, specified per stakeholder, are as follows:

\subsubsection{Content provider}

\begin{itemize}
\item \emph{Req. I:} Flexibility in choosing a license model.
\item \emph{Req. II:} Protection of the content (confidentiality). 
\item \emph{Req. III:} Enforcement of licenses.
\end{itemize}

\emph{Req. I}, in line with, for instance, \cite{music-licensing} for Spotify's music licensing, particularly asks for the following license models: 
\begin{itemize}
\item \emph{execute at most n-times}-models that allow only up to $n$ content executions.
\item \emph{pay per execute}-models that allow only one execution per payment. This is a special case of \emph{execute at most n-times}-license models.
\item \emph{execute until}-models that allow an unlimited number of executions of a content until a certain time.
\end{itemize}

It is important to note that the employed license model determines the way in which content is or can be used, which is why there is a need for freedom in choosing such models on the CP's side. One aspect CPs might want to define is, e.g., the number of times a user will be able to access content. Hence, the chosen license model needs to be capable of reflecting such statements. Employing different license models, applications like movie renting can be realized effortlessly. Further, degressive pricing can be established, meaning that users who buy a certain number of executions (in the \emph{execute at most n-times}-license model) are granted a discount: For example, if a user decides to buy $10$ instead of only one execution, he/she may be granted a $10 \%$ discount.

Regarding \emph{Req. II}, the CP---publicly providing the content in protected form---needs assurance that the content cannot be accessed by any party without proper rights. This requirement addresses third parties that have not purchased any content license. Moreover, \emph{Req. III} addresses users who are in possession of a license for the content they want to execute. The CP needs to be sure that the license terms will be checked before the content is allowed to be executed. This is necessary for, e.g., \emph{execute at most n-times}-models in which case the $n+1$\textsuperscript{st} execution shall not be allowed for users who have only paid for $n$ content executions.

\subsubsection{User}

\begin{itemize}
\item \emph{Req. IV:} Profile building (under a pseudonym) must not be possible for any involved party. To achieve \emph{Req. IV}, several requirements need to be met:
\end{itemize}

\begin{enumerate}
\item Anonymous content (license) buying towards content provider, and anonymous content execution towards content distributor. 
\item Unlinkability of content (license) purchases towards the content provider. 
\item Unlinkability of content executions towards the content distributor. 
\end{enumerate}

To precisely define \emph{anonymity} and \emph{unlinkability of items of interest} (IOIs) we have the following definitions \cite{unlinkability}:

\emph{Anonymity} and \emph{unlinkability of items of interest} (IOIs) are defined as~\cite{unlinkability}:

\begin{definition}[Anonymity]
Anonymity of a subject from an attacker's perspective means that the attacker cannot sufficiently identify the subject within a set of subjects, the anonymity set.
\end{definition}

\begin{definition}[Unlinkability delta]
The unlinkability delta of two or more IOIs from an attacker's perspective specifies the difference between the unlinkability of these IOIs taking into account the attacker's observations and the unlinkability of those IOIs given the attacker's a-priori knowledge only.
\end{definition}

\begin{definition}[Unlinkability of IOIs]
Unlinkability of IOIs is given iff the \emph{unlinkability delta} of two or more IOIs from an attacker's perspective is negligible. 
\end{definition}

Thus, (1) is achieved if neither the CP nor the CD can link a purchase/execution to a certain user, i.e., the user is not identifiable by those parties. Put~(2) in other words, after protocol execution, the CP's probability in linking content (license) purchases (IOIs) to each other has not changed compared to the probability in linking them to each other before the protocol execution. The same is true for (3), i.e., the CD's probability in linking content executions (IOIs) to each other does not change during protocol execution. 

Any party that was able to link IOIs to each other with non-negligible probability would be able to build a usage profile under a pseudonym. Such profile building under a pseudonym is an entity's ability to track a user's actions (under his/her pseudonym). If the CP/CD had such a profile, meaning it knew the accessed content, it might have a chance to de-anonymize the user, i.e., relate the pseudonym to an identity, given some external information---as shown, for example, by the de-anonymization of the ``anonymized'' \emph{Netflix} database \cite{netflix}. Thus, user privacy can only be achieved if profile building under a pseudonym can be prevented (unlinkability is a sufficient condition for anonymity \cite{unlinkability}).


\section{Privacy-preserving multiparty DRM system} \label{concept}

In a multiparty DRM system, there are multiple \emph{content providers} (CPs) that produce some sort of content, e.g., software, movies or music. The CPs only offer licenses for the given content. The content itself is delivered to the user by \emph{content distributors} (CDs), if and only if the user owns a valid license, that he/she has purchased from a CP before.

A multiparty DRM system, such as the one depicted in \refFig{fig:system_model} requires some questions to be answered: Without a TTP, it needs to be considered how license checking is performed and which entity is in charge of storing the user's purchased content. Moreover, as stated above, it is necessary to prevent the CP and the CD from creating user profiles (under a pseudonym).


To resolve these issues, we devised a protocol that involves a \emph{smartcard}. That is, instead of collaborating with a TTP, every user will employ a smartcard that has been provided to him/her. Such a smartcard contains a digital certificate that is used for (anonymous) authentication with CPs. Further, the user's smartcard assumes the role of his/her proxy. Thus, the user's smartcard executes an anonymous payment scheme (e.g., \cite{Tewari1998}) to pay the CP the corresponding amount for the requested license without disclosing any information about the user.

\subsection{Protocol Description}

The protocol is divided into several sub-protocols which work as follows.

\subsubsection{System Initialization} \label{system_initialization}

Let $\mathbb G_{1}$ and $\mathbb G_{2}$ be cyclic groups with the same prime order $q$, the security parameter $n = ||q||$, ${<}g{>}\ = \mathbb G_{1}$, and $Z = e(g,g) \in \mathbb G_{2}$.

Every smartcard is programmed and shipped by trustworthy\footnote{Note that these providers are not system-specific; any smartcard provider which is considered ``trustworthy'' today can be used to ship these smartcards.} smartcard providers. These providers install a private key $sk_{sc}$ and the corresponding digital certificate $cert_{sc}$ on every smartcard before shipping them. The private key and certificate are shared by all smartcards since they are used for anonymous authentication towards the CP during the process of purchasing content. Authentication of smartcards is required so that only legitimate smartcards can be used to purchase content, however, CPs must not be able to recognize smartcards. Moreover, the current time of production of the smartcard is set as the smartcard's timestamp $ts$.

Content offered by the CP is encrypted using a symmetric encryption algorithm such as \emph{AES} \cite{aes} and a separate content key $ck_{i}$ for each content $i$. 

The user sets up an anonymous payment scheme with his/her bank to get supplied with payment tokens $pt$.

\subsubsection{Content Purchase} \label{content_purchase}

\begin{figure*}[ht]
	\centering
	\includegraphics[width=\textwidth]{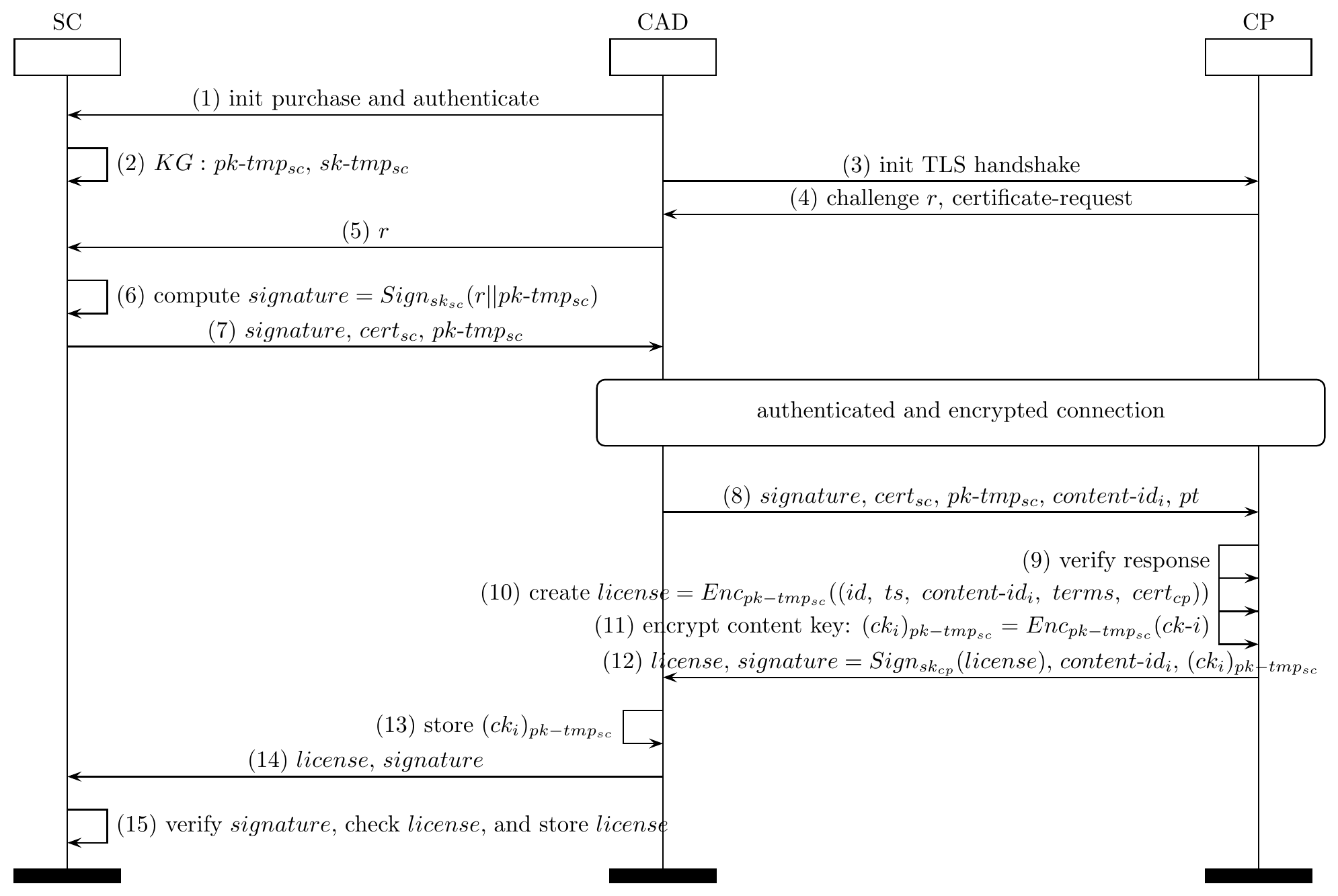}
	\caption{Content purchase protocol.}
	\label{fig:purchase}
\end{figure*}

The content purchase protocol is shown in \refFig{fig:purchase}. We assume that the connection between the user and the CP is anonymized (e.g., by using an anonymization network such as Tor \cite{tor}).

The user initiates the content purchase via his/her content access device (CAD), i.e., mobile phone, by authenticating towards the smartcard with his/her PIN (1) and initiating the TLS handshake with the CP (2). For reasons of readability, we do not show the entire TLS handshake \cite{tls_rfc}---using the RSA key exchange method---in the diagram, but rather focus on the important parts concerning the combination of the TLS handshake and the client authentication. In step (2), the smartcard executes the $KG$ algorithm as in \refSec{preliminaries} to generate a temporary key pair\footnote{A new temporary key pair is used for each content purchase.} $(pk$-$tmp_{sc} = (Z^{a_{1}}, g^{a_{2}}), sk$-$tmp_{sc} = (a_{1}, a_{2}))$, where $a_{1}, a_{2} \in \mathbb Z_{q}$ are chosen randomly. During the TLS handshake in step (4), the CP challenges the CAD's smartcard with a nonce\footnote{number used only once} $r$ and asks for the smartcard's certificate. The CAD forwards $r$ to the smartcard (5) which signs $r$ and $pk$-$tmp_{sc}$ with the smartcard's private key $sk_{sc}$ (6). The $signature$ and the smartcard's certificate $cert_{sc}$, as well as $pk$-$tmp_{sc}$ are forwarded to the CAD (7) and the CAD forwards them, together with the $content$-$id_{i}$ of the content $i$ to be bought, as well as the payment token $pt$ to pay for the license (8). From this moment on, the communication between CAD and CP is authenticated and encrypted via TLS. The CP verifies the response by checking the $signature$ (9). This way, the CAD's smartcard has anonymously authenticated towards the CP, meaning the CP knows that $pk$-$tmp_{sc}$ is from an authentic smartcard and the corresponding $sk$-$tmp_{sc}$ does not leave the smartcard. The CP creates the license for content $i$. This license includes a license identifier $id$, a timestamp $ts$, the $content$-$id_{i}$, the license $terms$, and the CP's certificate $cert_{cp}$ (10). Note that the license terms depend on the license model, i.e., for \emph{execute at most n-times}-models, the terms include the maximum number of allowed executions $n$. For other license models, the terms will differ accordingly. The license is encrypted under the smartcard's $pk$-$tmp_{sc}$. Moreover, the content key $ck_{i}$ for content $i$ is encrypted under $pk$-$tmp_{sc}$ as well (11). The $license$, the $signature$ of the license, the $content$-$id_{i}$ and the encrypted content key $(ck_i)_{pk-tmp_{sc}}$ are forwarded to the CAD (12). The CAD stores $(ck_i)_{pk-tmp_{sc}}$ (13) and forwards the $license$ and the $signature$ to the smartcard (14). The smartcard verifies the license's signature and decrypts the license with $sk$-$tmp_{sc}$. Then it checks whether the $id$ was not used before and whether $ts$ is newer than the current $ts$ on the smartcard---both to prevent replay attacks. The smartcard's $ts$ is then set to the newer $ts$ of the license.\footnote{Note that the smartcard does not have an internal clock and thus cannot keep track of (authenticated) time. The time can only be set via new and verified licenses.} Finally, the license is stored under the $content$-$id_{i}$ on the smartcard.~(15)

\subsubsection{Content Execution} \label{content_execution}

\begin{figure*}[ht]
	\centering
	\includegraphics[width=\textwidth]{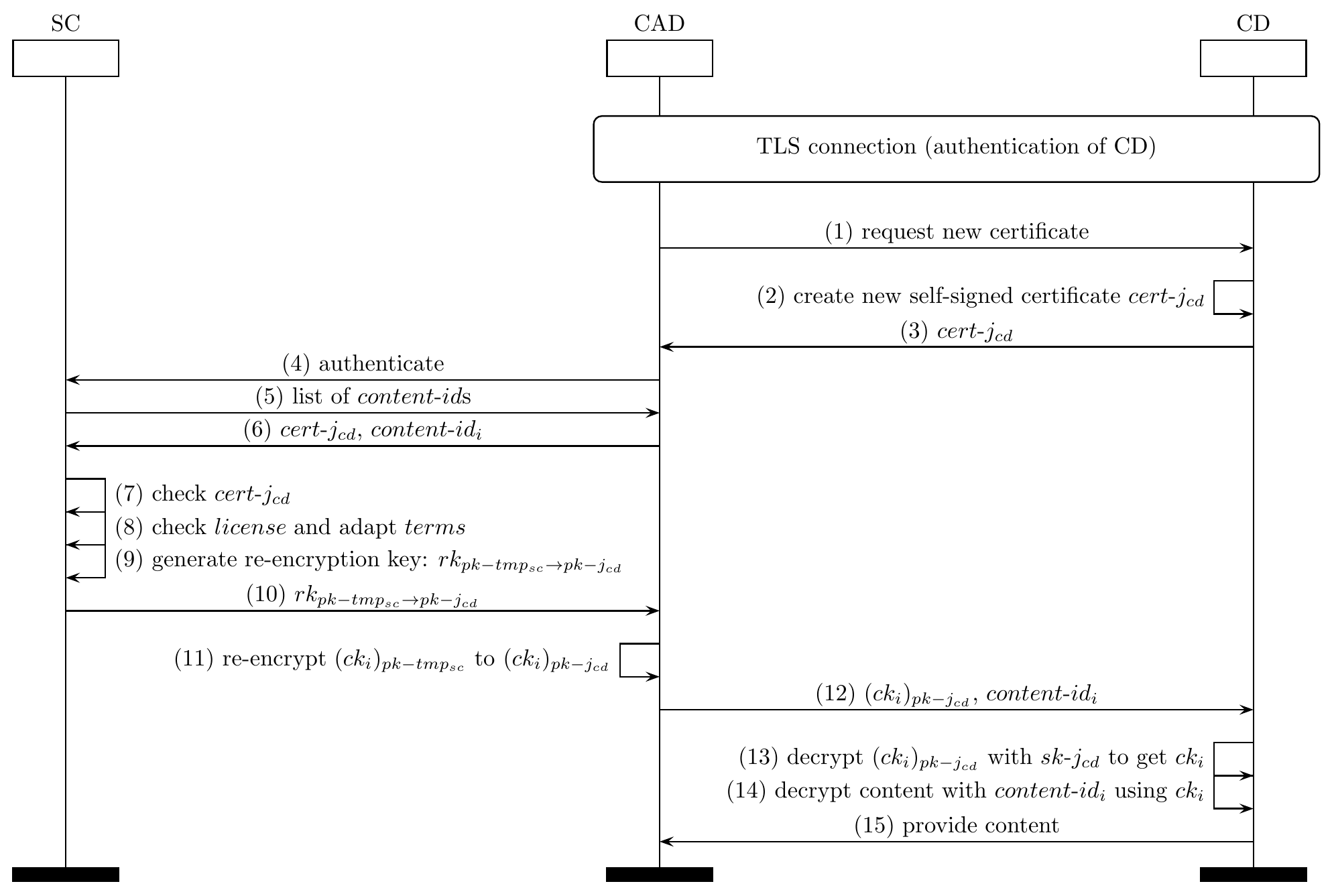}
	\caption{Content execution protocol.}
	\label{fig:execution}
\end{figure*}

The content execution protocol is shown in \refFig{fig:execution}. To playback the purchased content, the user first selects a CD of his/her choice (this choice could be automated as well, e.g., dependent of the region the user currently is in). We assume that the connection between the user and the CD is anonymized (e.g., by using an anonymization network such as Tor \cite{tor}).

The CAD establishes a TLS connection \cite{tls_rfc} with the CD---the CD authenticates towards the CAD with its certificate. The CAD afterwards requests a new certificate from the CD (1). The CD creates a new key-pair using $KG$ as in \refSec{preliminaries}: $(pk$-$j_{cd} = (Z^{a_{1}}, g^{a_{2}}), sk$-$j_{cd} = (a_{1}, a_{2}))$, where $a_{1}, a_{2} \in \mathbb Z_{q}$ are chosen randomly and $j$ denotes the $j$\textsuperscript{th} request to the CD. The $pk$-$j_{cd}$ is included in the newly generated certificate $cert$-$j_{cd}$, as well as a unique certificate $id$ and the current timestamp $ts$. The CD self-signs the certificate\footnote{The signing certificate was issued by a valid certificate authority, though.} (2). The certificate is forwarded to the CAD (3). The user authenticates towards the smartcard with his/her PIN entered on the CAD (4) and the smartcard then forwards the list of available $content$-$id$s, i.e. music/film titles, to the CAD (5). The user chooses the $content$-$id_{i}$ to be executed and forwards it, together with $cert$-$j_{cd}$ to the smartcard (6). The smartcard checks whether the signature of $cert$-$j_{cd}$ is valid, whether the CD was certified by a known certificate authority, whether the certificate $id$ was not used before and whether the $ts$ is newer than the current $ts$ on the smartcard. If these tests pass, the new $ts$ from the certificate is set on the smartcard (7). It is important to note, that the smartcard checks whether the certificate really belongs to a CD. If this was not the case, the user might be able to launch an attack by including a self-signed certificate that he/she has generated himself/herself. Hence, if the smartcard would not verify that the certificate belonged to a CD, the user might acquire a re-encryption key from the smartcard that allowed him/her to decrypt the content key, granting him/her unlimited access to the content. Furthermore, the smartcard checks whether the license $terms$ still allow the content to be played back, i.e., whether the number of executions is $<\!n$, the current $ts$ is before the end date of the license, etc. If this is the case, the $terms$ are updated, that is, e.g., a counter that counts the number of executions is incremented by one for an \emph{execute at most n-times}-license (8). Then, the smartcard generates the re-encryption key $rk_{pk-tmp_{sc} \rightarrow pk-j_{cd}}$ by using the $RG$ algorithm as in \refSec{preliminaries}, taking as input the CD's public key $g^{a_{2}} \in pk$-$j_{cd}$, and its own private key $a_{1} \in sk$-$tmp_{sc}$ (as created during the content purchase described in \refSec{content_purchase}) (9). The re-encryption key is then forwarded to the CAD~(10). The CAD re-encrypts the encrypted content key $(ck_i)_{pk-tmp_{sc}}$ by employing the $R$ algorithm as in \refSec{preliminaries} with $rk_{pk-tmp_{sc} \rightarrow pk-j_{cd}}$ as input to retrieve $(ck_i)_{pk-j_{cd}}$---i.e., the encrypted content key under the CD's public key (11). The re-encrypted content key is then forwarded to the CD (12) and the CD decrypts the ciphertext using the $D$ algorithm as in \refSec{preliminaries} with its private key $a_{2} \in sk$-$j_{cd}$ as input to retrieve $ck_{i}$ (13). The content---retrieved from the CP---can now be decrypted by the CD using $ck_{i}$ and the symmetric scheme as employed during the system initialization phase (\refSec{system_initialization}) (14). Eventually, the content is provided, for example, streamed, to the user's CAD (15).

\subsubsection{Authorization Categories} \label{authorization_categories}

Similar to \emph{``Authorization Categories''} \cite{perlman_privacy_preserving}, we came up with the following addition to our protocol: There might be content that should not be accessible to everybody, such as X-rated content. To be able to check whether a user should be allowed to access certain content, we employ the smartcard of the user's CAD. Before initially obtaining a smartcard, the user provides certain information to the smartcard provider (e.g., his/her passport or state ID). The smartcard provider will then securely\footnote{Secure storage in this context especially means integrity-protection; this can be achieved by a signature calculated by the smartcard provider.} store the required information on the user's smartcard. Note that the smartcard, as well as the smartcard provider may obtain such information without invading the user's privacy.

If we assume that the user's smartcard now contains information like the user's date of birth or home country, it can easily check whether or not the user is allowed to access the content he/she wishes. This means that if the user requests access to, for instance, X-rated content, the smartcard will check the user's date of birth and according to this information either allow or deny access to the queried content (that is, if the access is denied, the smartcard will not carry out the content purchase as described in \refSec{content_purchase}).

Compared to the three approaches described in \cite{perlman_privacy_preserving}, we do not need to employ complicated protocols for checking a user's authorizations, since the required information are stored and checked by the user's smartcard only (i.e., neither CP nor CD are involved in the checking for authorization). Hence, none of the two, CP and CD, are burdened with additional computations. Due to the fact that the smartcard is assumed to be a tamper-resistant device, the user cannot manipulate the information stored on the smartcard and therefore is unable to access content that he/she lacks the authorization to. Further, this supplement to our protocol is optional and can be employed according to certain policies (e.g., in some countries particular checks might be required by law, whereas other countries do not pose any such requirements).

\section{Evaluation and Discussion} \label{evaluation}

In this section we discuss the protocol's performance, determine whether it meets the posed requirements and compare our approach to related work.

\subsection{Performance Analysis} \label{performance_analysis}

%
%

The protocol requires the user's CAD to execute the most complex tasks, i.e., perform the re-encryption of the content key. Apart from the re-encryption, the content provider and the smartcard are involved in a challenge-response protocol for authentication of the smartcard which is not too expensive. Further, the CP has to encrypt the content key using the smartcard's public key and the content using the content key. The latter is a symmetric encryption that has to be executed only once per content (and not for every purchase/execution). Additionally, the content distributor has to decrypt the re-encrypted content key as well as the content obtained from the CP. The required generation of keys is not expensive. The only concern is the re-encryption key generation that is performed on the user's smartcard. We were able to show that current smartphones are easily capable of executing the required tasks by implementing a demo application on an Android smartphone. We have implemented the re-encryption using the jPBC (\emph{Java Pairing Based Cryptography}) library\footnote{\url{http://gas.dia.unisa.it/projects/jpbc/}}. The app that has been developed re-encrypts $128\,$Bytes of data---the length of a symmetric key to be encrypted---in $302\,ms$ on a Samsung Galaxy Nexus ($2 \times 1.5\,GHz$) running Android 4.2. Due to a lack of a proper smartcard\footnote{According to the specifications, the NXP JCOP card 4.1, V2.2.1 can be used to implement the needed functionality.}, we could not implement the re-encryption key generation algorithm $RG$ as in \refSec{preliminaries}. Thus, to show the practicability of the implementation, we must refer to \cite{smartcard}. The authors have implemented elliptic curve scalar point multiplications and additions for a smartcard in C and Assembler---which are needed in our approach as well. As the authors conclude, the standard Javacard API (version 2.2.2) cannot be used as the available EC Diffie-Hellman key exchange only provides the hashed version of the key derivation function. \cite{smartcard} However, we need the immediate result of the key derivation function, i.e., the result of the EC point multiplication. Our own implementation of the EC point multiplication on the smartcard's CPU did not yield practicable results---as the efficient cryptographic co-processors could not be utilized due to proprietary code.

\subsection{Evaluation of Requirements}


\emph{Req. I: Flexibility in choosing a license model: } The CP is able to provide different kinds of rights to users for content playback---users pay the corresponding prices. Our system is flexible enough to allow for the most popular models like \emph{flatrate}, \emph{execute at most} $n$\emph{-times}, \emph{execute until a certain date}, etc.

\emph{Req. II: Protection of the content: } The CP distributes its content only in encrypted form, as described in \refSec{system_initialization}. Thus, none of the parties not in possession of the content decryption key is able to access the content. 

\emph{Req. III: Enforcement of licenses: } Smartcards, as trusted devices, are used in our protocol to enforce licenses. Thus, if the smartcard's check of a license fails, the proper re-encryption key is not generated and the user is not able to execute the content. A replay attack of the user with an ``old'' CD certificate will fail as the smartcard will not accept the $ts$---since it is older than the current one stored on the smartcard. The smartcard's property of tamper-resistance is required since we assumed users to be active adversaries in \refSec{system_model}. 

Note that during the protocol execution, the re-encryption key is ``leaked''~\cite{re-encryption_key_leakage} to the untrusted CAD\footnote{Note that the CAD is in the user's domain and the user is seen as an attacker who tries to circumvent the license checking.} in step (10). As a consequence, without protection mechanisms, a user who has obtained a re-encryption key from his/her smartcard that was valid for a certain CD would be able to re-encrypt a content key as many times as he/she desired for this CD. This would defeat the purpose of license checking, since the user could playback content in an unrestricted fashion. This is why we have devised a method to prevent the user from reusing re-encryption keys: The transience of the CD's certificates and thus, public keys (2), prevents replay attacks, meaning that the user cannot reuse the re-encryption key later on; if the user re-encrypted a content key under a temporary public key of the CD that was already used before, the CD would reject the request of content execution, because the temporary public key has already been invalidated after a former execution. 
Hence, the user relies on his/her smartcard to generate a valid re-encryption key which it will only do if the user's license for the corresponding content is (still) valid. \textsc{Ateniese et al.} \cite{ateniese_improved} propose an addition to the proxy re-encryption scheme used in this paper: if a trusted server broadcasting some random number for each time period was assumed, the proxy re-encryption could be adapted so that only temporary re-encryptions are possible. Thus, the ``leakage'' problem could be minimized. However, in our scenario we do not want to introduce such a TTP and moreover, as stated above, the smartcard does not have an internal clock---time periods would introduce additional challenges. 

\begin{figure}[ht]
	\centering	\includegraphics[width=0.55\columnwidth]{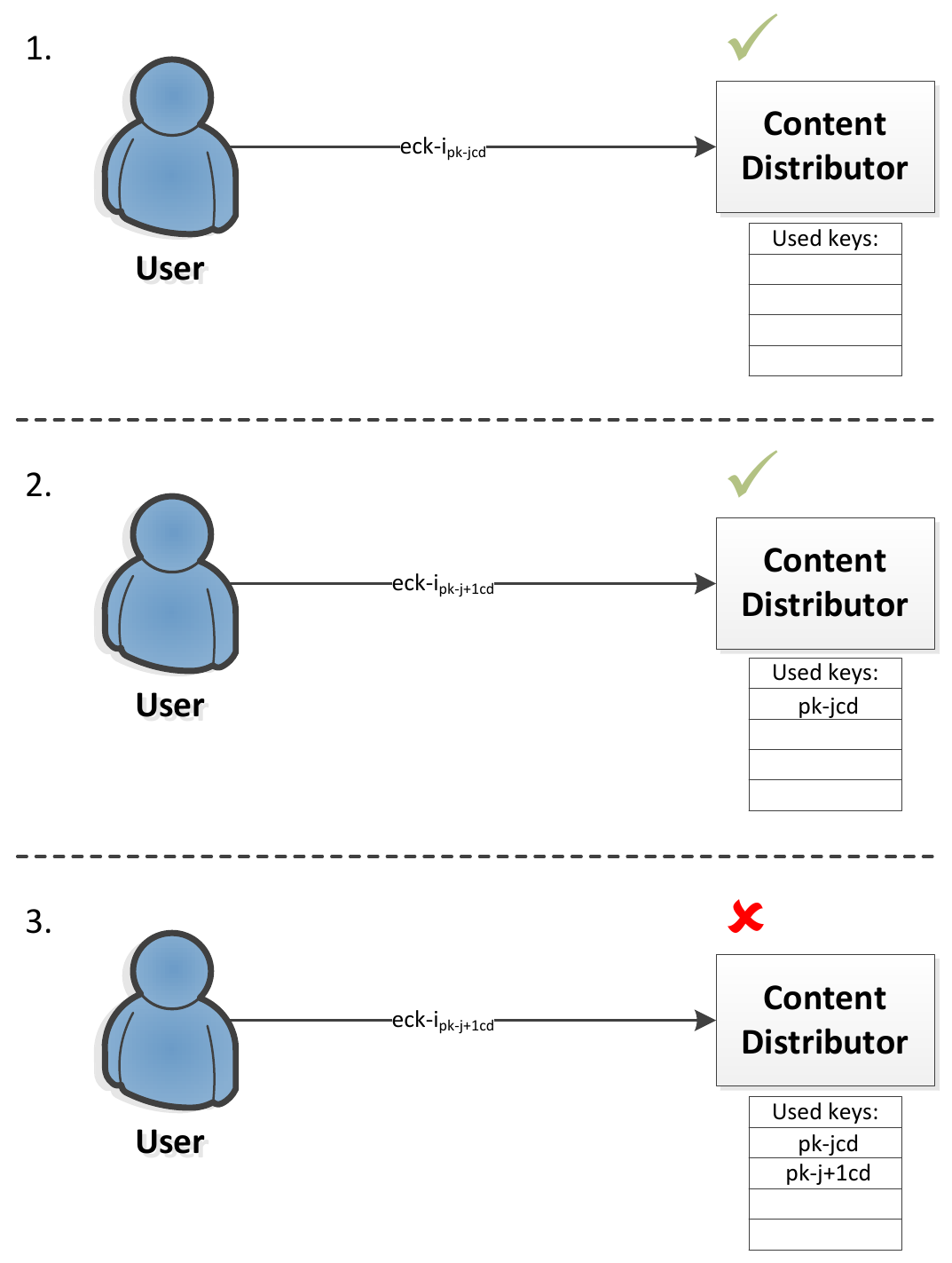}
	\caption{Prevention of re-encryption key leakage in our protocol.}
	\label{fig:re-encryption_key_leakage_prevention}
\end{figure}

\emph{Req. IV: Profile building (under a pseudonym) must not be possible for any party involved in the protocol: } As required in \refSec{requirements}, the following points need to be met to prevent profile building under a pseudonym: 

\begin{enumerate}
\item Anonymous content (license) buying towards content provider, and anonymous content execution towards content distributor.
\item Unlinkability of content (license) purchases towards the content provider.
\item Unlinkability of content executions towards the content distributor.
\end{enumerate}

(1) As we have pointed out in \refSec{content_purchase}, users anonymously pay for content (licenses), i.e., they do not need to register with the CP/CD and need not provide their payment details, which is why they stay anonymous during their transactions with CP and CD.

(2) All the smartcards use the same certificate for anonymous authentication towards the CP (\refSec{content_purchase}), thus the CP cannot link different purchases made with the same smartcard. The smartcard's public key $pk$-$tmp_{sc}$ is newly generated for each content (license) purchase (\refSec{system_initialization})---preventing the CP from linking purchases to each other. Moreover, the used anonymous payment scheme \cite{Tewari1998} provides unlinkability of individual payments. Furthermore, we assumed the connection between user and CP to be anonymized via Tor. Thus, unlinkability of content (license) purchases is achieved. 

(3) The user only provides the re-encrypted content key to the CD. Content $i$ is only encrypted once during the initialization phase with $ck_{i}$ and thus, $ck_{i}$ does not contain any information connected to the user or the user's CAD. As a new re-encryption key is generated for each content execution, the encrypted content key ``looks'' different for the CD each time and hence, the CD cannot link any pair $(ck_i)_{pk-j_{cd}}$, $(ck_i)_{pk-k_{cd}}$, for $j \neq k$ to each other (see \refFig{fig:re-encryption_key_leakage_prevention}). Further, we assumed the connection between the user and the CD to be anonymized via Tor. Therefore, multiple transactions executed by the user are unlinkable for the CD.

Moreover, even if an attacker gets access to the user's mobile phone, he/she does not learn which content has been bought and executed. The list of available content is only revealed by the smartcard after authentication with the proper PIN and the mobile phone application does not keep track of executed content. 

Thus, to sum it up, profile building (even under a pseudonym) is neither possible for the CP nor the CD.

\subsection{Comparison to related work} \label{comparison}

In \refTable{tab:comparison} we compare our proposed scheme to related work in the field of privacy-preserving digital rights management. 


\begin{table*}[ht]
	\small
	\centering
	\caption{Comparison of our scheme to related work in terms of properties.}
	\label{tab:comparison}
	\renewcommand{\arraystretch}{1.25}
	\renewcommand{\tabcolsep}{0.15cm}
	\begin{tabular}{@{}p{0.275\textwidth}p{0.12\textwidth}p{0.12\textwidth}p{0.12\textwidth}p{0.12\textwidth}p{0.12\textwidth}p{0.12\textwidth}@{}}
		\toprule[.15em]
		\textsc{Properties} & \textbf{Paper\newline at hand} & \textbf{\cite{practical}} & \textbf{\cite{proxy}} & \textbf{\cite{multilevel}} & \textbf{\cite{nottp}}\\ 
		\midrule
		\textbf{Need for TTP} & no & yes & yes & yes & no\\
		\midrule
		\textbf{Need for trusted\newline hardware} & yes & no & no & no & yes\\
		\midrule
		\textbf{Flexibility in\newline choosing a\newline license model} & yes  & yes  & yes & no & yes\\
		\midrule
		\textbf{Unlinkability of\newline content executions} & yes & yes & yes & yes & no\\
		\midrule
		\textbf{Computational\newline efficiency} & good & medium & bad & good & good\\
		\midrule
		\textbf{Flexibility in\newline choosing content\newline distributor} & yes & yes & yes & yes & yes\\
		\bottomrule[.15em]
	\end{tabular}
\end{table*}

\vspace{-0.5cm}

\subsubsection{Need for TTP}

One of the main advantages of our scheme compared to related work is that it does not need a trusted third party which is involved in the license checking process as in \cite{proxy, practical} during each content execution. In~\cite{multilevel}, the license server constitutes the TTP. However, it is not involved in the protocol for each single content execution but only once, when retrieving the license. 


\subsubsection{Need for trusted hardware}

In our protocol a smartcard performs the license checking. Trusted hardware is not needed by other protocols that rely on some TTP. A trusted platform module (TPM) is needed in the protocol presented in \cite{nottp} to securely store tokens at the user's computing platform. 

\subsubsection{Flexibility in choosing a license model}

The protocols presented here and in \cite{proxy, practical} allow for a license model to be chosen freely, e.g., content execution at most $n$ times, up to a certain point of time, etc. The protocol presented in \cite{multilevel} does not allow such flexibility---once a license is bought for some content, it may be executed by the user as often as desired. The authors of \cite{nottp} do not clearly state whether differentiated license models are intended. From the protocol's point of view, it should be possible to implement, e.g., \emph{execute at most} $n$ \emph{times}-models as a token set provided by the content owner. Such token sets could include $n$ tokens. Further, licenses that allow only a single content execution could be mapped to each token by the content provider\footnote{Content distributor in our scenario.} later on.

\subsubsection{Unlinkability of content executions}

All of the approaches covered here, except for \cite{nottp}, provide unlinkability of content executions and thus, prevent any party from building a content usage profile (under a pseudonym).


\subsubsection{Computational efficiency}

In terms of computational overhead, our proposed scheme is very efficient, as shown in \refSec{performance_analysis}. The scheme presented in \cite{practical} makes use of a number of different cryptographic primitives and thus performs less well. In \cite{proxy}, the entire content is re-encrypted for each content execution. Efficient standard cryptographic primitives are used in \cite{multilevel, nottp}. 

%
%


\subsubsection{Flexibility in choosing content distributor}

All the schemes presented in this overview provide users with the possibility to freely choose the CDs. In other two-party DRM scenarios, such a flexibility is typically not provided. 


\section{Conclusion} \label{conclusion_outlook} 


We have come up with a privacy-preserving multiparty DRM concept. Users can anonymously buy content licenses from a content provider and anonymously execute the content at any content distributor by, for example, streaming the content from content distributors nearby. Anonymity in this context means that none of the involved parties is able to build a content usage profile---not even under a pseudonym. In contrast to related work on privacy-preserving DRM, the approach presented in this paper does not employ a trusted third party. Smartcards are used to check content licenses and grant execution if the license still allows the respective content to be executed---thus, enabling very differentiated license models. We implemented our concept on a state-of-the-art smartphone and proved its practicability for a multiparty DRM scenario in a mobile environment in which a user buys a license allowing the playback of, e.g., some TV show---roaming in different regions, the user is free to choose the nearest streaming server (content distributor) and hence, getting the best throughput. As licenses are bound to the user's smartcard, content usage is device-independent and the user may use any of his/her devices to playback the content. 

\bibliography{Literatur/literatur}
\bibliographystyle{splncs} 

\end{document}